\documentclass[fleqn,usenatbib]{mnras}

\usepackage{newtxtext,newtxmath}

\usepackage[T1]{fontenc}
\usepackage{xcolor}

\DeclareRobustCommand{\VAN}[3]{#2}
\let\VANthebibliography\thebibliography
\def\thebibliography{\DeclareRobustCommand{\VAN}[3]{##3}\VANthebibliography}
\newcommand{\ag}[1]{{\color{teal}{[{\bf AG:} #1}]}}

\usepackage{graphicx}	
\usepackage{amsmath}	

\usepackage{amssymb}	
\usepackage{comment}

\title[Reionization inference from kSZ]{Inferring reionization and galaxy properties from the patchy kinetic Sunyaev-Zel'dovich signal}

\author[I. Nikolić et al.]{
Ivan Nikolić,$^{1}$\thanks{E-mail: ivan.nikolic@sns.it}
Andrei Mesinger,$^{1}$
Yuxiang Qin$^{2,3}$
and Adélie Gorce$^{4}$
\\
$^{1}$Scuola Normale Superiore, Piazza dei Cavalieri 7, 56126 Pisa, PI, Italy\\
$^{2}$School of Physics, University of Melbourne, Parkville, VIC 3010, Australia\\
$^{3}$ARC Centre of Excellence for All Sky Astrophysics in 3 Dimensions (ASTRO 3D)\\
$^4$Department of Physics and Trottier Space Institute, McGill University, Montreal, QC H3A 2T8, Canada
}

\date{Accepted XXX. Received YYY; in original form ZZZ}

\pubyear{2022}

\begin{document}
\label{firstpage}
\pagerange{\pageref{firstpage}--\pageref{lastpage}}
\maketitle

\begin{abstract}
The patchy kinetic Sunyaev-Zel’dovich (kSZ) signal is an integral probe of the timing and morphology of the epoch of reionization (EoR).  Recent observations  have claimed a low signal-to-noise (S/N) measurement, with a dramatic increase in S/N expected in the near future. In this work, we quantify what we can learn about the EoR from the kSZ signal.  We perform Bayesian inference by sampling galaxy properties and using forward-models of the kSZ as well as other EoR and galaxy observations in the likelihood. Including the recent kSZ measurement obtained by the South Pole Telescope ($\mathcal{D}_{3000}^{\textrm{pkSZ}} = 1.1_{-0.7}^{+1.1} \mu$K$^2$) shifts the posterior distribution in favor of faster and later reionization models, resulting in lower values of the optical depth to the CMB: $\tau_e = 0.052_{-0.008}^{+0.009}$ with a 68\% confidence interval (C.I.).
The combined EoR and UV luminosity function observations also
imply a typical ionizing escape fraction of $0.04_{-0.03}^{+0.05}$ (95\% C.I.), without a strong dependence on halo mass.
We show how the patchy kSZ power from our posterior depends on the midpoint and duration of reionization: a popular parametrization of EoR timing. For a given midpoint and duration, the EoR morphology only has a few percent impact on the patchy kSZ power in our posterior.  However, a physical model is needed to obtain tight constraints from the current low S/N patchy kSZ measurement, as it allows us to take advantage of complimentary high-$z$ observations.
Future high S/N detections of the patchy kSZ should decrease the current uncertainties on the timing of the EoR by factors of $\sim$2 -- 3.
\end{abstract}

\begin{keywords}
cosmology: cosmic background radiation -- dark ages, reionization, first stars -- diffuse radiation -- large-scale structure of Universe -- early Universe -- galaxies: high-redshift
\end{keywords}



\section{Introduction}

The epoch of reionization (EoR) is a major milestone in the Universe's evolution. Although many questions remain, recent years have seen a dramatic increase in the volume of data available to probe the cosmological frontier. These include: (i) high-redshift QSO spectra \citep[e.g.][]{Bolton2010, Becker15, Irsic17, Bosman2018, Gaikwad20}; (ii) Lyman alpha emitting galaxies \citep[e.g.][]{Stark2010,Stark17, Konno18, Hoag19, Mason2019,Leonova2022MNRAS.515.5790L,Endsley2022MNRAS.517.5642E}; (iii) the optical depth to the CMB \citep[e.g.][]{PlanckCollaborationAdam_2016, Planck2020, Heinrich21}; (iv) UV luminosity functions \citep[LF, e.g.][]{Finkelstein15, Bouwens2015, Bouwens2016, Oesch18, Bhatawdekar19,Bouwens2023MNRAS.523.1036B,Harikane2023ApJS..265....5H}; (v) preliminary upper limits on the 21\,cm power spectrum \citep{Mertens20, Trott20, HERA22, HERA23}. This trend is set to culminate in the coming decade with 21-cm maps of the first billion years from the Square Kilometre Array (SKA)\footnote{\url{https://www.skatelescope.org/}}.

A complementary probe that has arguably seen less attention is provided by the patchy kinetic Sunyaev-Zel'dovich (kSZ) signal.  The kSZ is sourced by the Doppler shifting of CMB photons that scatter off of free electrons, resulting in secondary temperature anisotropies.  It is typically separated into post-EoR (or homogeneous) and EoR (or patchy) contributions.  The patchy kSZ is determined by the timing, duration and morphology of the EoR.  Thus, measuring its shape and amplitude could inform us about the evolution of this cosmic milestone as well as the galaxies that sourced it \citep[e.g.][]{Iliev2007, Mesinger2012, Park2013, Choudhury2021, Begin2022}.

Measurements of the patchy kSZ have historically focused on  the angular multipole $l=3000$ (roughly corresponding to 4~arcmin, or a comoving scale of 20\,Mpc during the EoR).  At lower multipoles the primary CMB anisotropies are increasingly dominant, while at higher multipoles systematics such as the cross-correlation between the thermal Sunyaev-Zel'dovich (tSZ) and dusty galaxies become even more challenging.  The two telescopes actively targeting the kSZ, the Atacama Cosmology Telescope (ACT) and the South Pole Telescope (SPT), have until recently only published upper limits \citep{Lueker2010, Das2011, Shirokoff2011, Reichardt2012, Dunkley2013,Das2014,  George2015}. Strong foregrounds, including bright extragalactic sources, 
as well as modelling uncertainties remain very challenging. However, the SPT collaboration recently claimed a low signal to noise (S/N) measurement of the patchy kSZ signal: $\mathcal{D}_{3000}^{\textrm{pkSZ}} = 1.1_{-0.7}^{+1.1}\,\mu$K$^2$ \citep[68\% C.I.,][]{Reichardt2021}. 
These relatively low values qualitatively point to a much later and more rapid EoR compared to original estimates \citep[e.g.][]{Mcquinn2005, Zahn2005, Iliev2007, Zahn2012, Mesinger2012,  Park2013, Calabrese2014, Alvarez2016, Paul2021}. Future telescopes, such as the Simons Observatory\footnote{\url{https://simonsobservatory.org/}} \citep{Ade2019}, CMB-Stage~4\footnote{\url{https://cmb-s4.org/}} \citep{Abitbol2017} and  CMB-HD\footnote{\url{https://cmb-hd.org/}} \citep{Sehgal2019}, should help better characterize the CMB foregrounds and related systematics to narrow down error bars.

However, interpreting a tentative detection of the kSZ is difficult.  Firstly, one needs to statistically separate the homogeneous and patchy contributions from the total kSZ power. Secondly, the patchy kSZ power is an integral measurement of the EoR, and as such is prone to strong astrophysical parameter degeneracies.  Robust interpretation therefore must rely on additional, complementary observations of the EoR and high-redshift galaxies.

Here we quantify what we can learn from the recent kSZ measurement using a fully Bayesian framework.  Unlike previous works, we directly sample empirical properties of galaxies that drive the EoR, creating 3D lightcones on-the-fly. This allows us to: (i) self-consistently sample different EoR morphologies when comparing against kSZ observations (instead of the common approach of fixing the morphology and empirically varying the midpoint and duration of the EoR); (ii) combine independent high-$z$ galaxy and EoR observations when computing the posterior; and (iii) set physically-meaningful priors.

This paper is organized as follows. In Sec.~\ref{section2} we discuss how we compute the patchy kSZ signal. Our Bayesian framework, combining the kSZ with complementary observations, is summarized in Sec.~\ref{sec:complimentary_observations}. We present and discuss our results in Sec.~\ref{results}. In Sec.~\ref{self:self-consistent} we quantify how accurately the midpoint and duration of reionization can predict the patchy kSZ at $l=3000$.  Finally, we conclude in Sec.~\ref{conclusion}.
Throughout this work, we assume standard $\Lambda$CDM cosmological parameters ($\Omega_\textrm{m}, \Omega_\textrm{b}, \Omega_{\Lambda}, h, \sigma_8, n_\textrm{s} = 0.321, 0.049, 0.679, 0.67, 0.81, 0.963$), consistent with the latest estimates from \citet{Planck2020}.  Unless stated otherwise, we quote all quantities in comoving units.

\section{The patchy kinetic Sunyaev-Zel'dovich signal}
\label{section2}

The secondary temperature anisotropy of the CMB due to the kinetic Sunyaev-Zel'dovich effect in the line of sight (LoS) direction $\mathbf{\hat{u}}$ can be written as:
\begin{equation}
\begin{aligned}
    \delta T_{\textrm{kSZ}} & = \frac{\Delta T}{T} \left(\mathbf{\hat{u}}\right)\\ & =  \sigma_{\rm T} \int \textrm{d}z \ \left(\frac{\textrm{d}t}{\textrm{d}z}\right) \ e^{-\tau_e(z)} n_e \ \mathbf{\hat{u}} \cdot \mathbf{v}  \\ & = \sigma_{\rm T} \int \textrm{d}z \ \left(\frac{\textrm{d}t}{\textrm{d}z}\right) \ e^{-\tau_e(z)} \ x_e \ n_b \ \mathbf{\hat{u}} \cdot \mathbf{v}.
    \label{general_equation}
\end{aligned}
\end{equation}
Here, $\sigma_\textrm{T}$ is the Thomson scattering cross-section, $n_e$ is the number density of electrons\footnote{Here we assume helium is doubly ionized at $z<3$, and singly ionized at the same fraction as hydrogen during the EoR.} which can be expanded as the product of the ionized fraction ($x_e$) and baryon density ($n_b$), $\mathbf{v}$ is the velocity of electrons,  and $\tau_e$ is the optical depth of CMB photons up to redshift $z$:
\begin{equation}
    \tau_e (z) = \sigma_{\rm T} \int_0^z \textrm{d}z' \ c \frac{ \textrm{d}t}{\textrm{d}z'} \, n_e .
\end{equation}

The redshift integral in equation~\eqref{general_equation} is generally separated into a post-reionization (or homogeneous) component and one due to patchy reionization.  Observations measure the total kSZ power spectrum\footnote{The power spectrum is defined as $\mathcal{D}_l^{\textrm{kSZ}} = \frac{l(l+1)}{2\pi}C_l^{\textrm{kSZ}}$, where $C_l^{\textrm{kSZ}} = T_{\textrm{CMB}}^2 |\overline{\delta T}_{\textrm{kSZ}} (k)|^2$, $T_{\textrm{CMB}}$ is the mean CMB temperature and $\overline{\delta T}_{\textrm{kSZ}} (k)$ is the Fourier transform of $\delta T_{\textrm{kSZ}}$.}, which is the sum coming from these two respective components: $\mathcal{D}_l^{\textrm{kSZ}} = \mathcal{D}_l^{\textrm{hkSZ}} + \mathcal{D}_l^{\textrm{pkSZ}}$. The post-reionization kSZ power spectrum, $\mathcal{D}_l^{\textrm{hkSZ}}$, is dominated by fluctuations in $n_b$ during the era of cluster formation at $z\lesssim1$ (e.g. \citealt{Shaw2012}), while the patchy kSZ power spectrum, $\mathcal{D}_l^{\textrm{pkSZ}}$,  is dominated by order unity fluctuations in $x_e$ during the EoR at $z\gtrsim5$ (e.g. \citealt{Alvarez2016}).  Constraining the patchy kSZ thus requires statistically accounting for the post-EoR (homogeneous) kSZ signal;  we summarize how this was done for recent observations in Sec.~\ref{observing_section}. Because we do not know \textit{a priori} the reionization redshift, here we {\it define} the patchy kSZ component as the contribution to equation~\eqref{general_equation} of redshifts above $z \geq 5$.  We note that this is a lower value compared to some previous choices in the literature.  It is motivated by recent Lyman alpha forest data whose interpretation requires a late reionization, ending at $5.3 \lesssim z \lesssim 5.6$ (\citealt{Qin2021, Choudhury2021b, Bosman2022}, Qin et al. in prep).

The kSZ power is typically measured at $l=3000$; smaller multipoles are increasingly dominated by primary CMB anisotropies, while larger multipoles become swamped by other foregrounds such as dusty galaxies \citep[e.g.][]{Zahn2012, Alvarez2016, Reichardt2012}. During the EoR, $l=3000$ roughly corresponds to physical scales of $\sim20$\,cMpc. Therefore, measurements of the patchy kSZ at this multipole are sensitive to the EoR morphology on these scales, as well as the timing and duration of the corresponding epochs. Simulation box sizes larger than about 300\,cMpc are sufficient to capture the ionization power spectra on those scales \citep[e.g.][]{Iliev2014, Harman2020}. Unfortunately, the kSZ is determined to leading order by the velocity-ionization cross power, and much larger scales (above 1 cGpc) are required to capture the fluctuations in the velocity field and corresponding velocity-ionization cross-power at $l\sim3000$ \citep[e.g.][]{Shaw2012, Alvarez2016}.  Given that radiative transfer simulations on such large scales are computationaly prohibitive, more approximate schemes are required to calculate the patchy kSZ signal.

The patchy kSZ power is sometimes computed analytically (with some terms calibrated to smaller numerical simulations; e.g. \citealt{Park2013, Gorce2020}) but at the price of neglecting the contribution of higher order correlations (above two points) which can represent up to $10\%$ of the total patchy power \citep{Alvarez2016}.  More importantly, it is difficult to associate prior probabilities on the "effective" parameters of such models; priors are important for inference from a low S/N detection whose likelihood is not strongly constraining.
 Instead, in this work, we choose to compute the patchy kSZ signal by ray-tracing through large 3D lightcone simulations with approximate radiative transfer (so-called semi-numerical simulations; e.g. \citealt{Zahn2012, Mesinger2012, Battaglia2013, SeilerHutter_2019, Choudhury2021, Chen2022}).  Our self-consistent approach allows us to incorporate multi-frequency observations of the EoR and high-$z$ galaxies in the likelihood.  We discuss how this is done in the following section.

\subsection{Computing the patchy kSZ from galaxy-driven EoR simulations}
\label{estimating_section}

In this work we extend the public simulation package {\tt 21cmFAST}\footnote{\url{https://github.com/21cmfast/21cmFAST}} (e.g. \citealt{2007ApJ...669..663M, 2011MNRAS.411..955M, 2020JOSS....5.2582M}) to forward-model the patchy kSZ signal together with other observables. {\tt 21cmFAST} is a semi-numerical code used for generating cosmological simulations of the early Universe. It computes the evolved density and velocity fields using second-order Lagrangian perturbation theory (e.g. \citealt{Scoccimarro1998}). The ionization field is generated from the density field by comparing the cumulative number of ionizing photons produced by galaxies to the number of hydrogen atoms plus cumulative number of IGM recombinations, in spherical regions with decreasing radii, $R$ \citep[e.g.][]{Furlanetto2004}.  Specifically, a cell is marked as ionized if at any radius:
\begin{equation}
    n_\textrm{ion} \ge (1 + n_{\textrm{rec}} )(1 - x_e ),
\end{equation}
where $n_{\textrm{rec}}$ is the cumulative number of recombinations per baryon computed according to the sub-grid scheme of \citet{Sobacchi2014}, $x_e$ accounts for pre-ionization by X-ray photons, and $n_{\textrm{ion}}$ is the cumulative number of ionizing photons per baryon, with quantities averaged over the sphere of radius $R$:
\begin{equation}
    n_{\textrm{ion}} = \frac{1}{\rho_b} \int_0^{\infty} \textrm{d} M_{\rm h} \ \frac{\textrm{d}n (M_{\rm h}, z | R, \delta_R)}{\textrm{d} M_{\rm h}} \ f_{\textrm{duty}} \ M_{\ast} \ f_{\textrm{esc}} \ N_{\gamma/b}.
    \label{eq:nion}
\end{equation}
Here $dn/dM_{\rm h}$ is the conditional halo mass function, $N_{\gamma/b}$ is the number of ionizing photons per stellar baryon, $f_{\textrm{esc}}$ is the escape fraction of ionizing photons, and $f_{\textrm{duty}}$ corresponds to the fraction of halos that host star forming galaxies.  

Here we adopt the flexible parameterization from \citet{Park2019}.  Specifically, $f_{\textrm{duty}}$ decreases exponentially below a characteristic mass scale, $M_{\textrm{turn}}$, due to inefficient gas cooling and/or feedback (e.g. \citealt{Sobacchi2013a, Xu2016, Mutch2016}):
\begin{equation}
    f_{\textrm{duty}}(M_{\rm h}) = \exp \left(-\frac{M_{\rm h}}{M_{\textrm{turn}}}\right).
    \label{eq:fduty}
\end{equation}
The ionizing escape fraction $f_{\textrm{esc}}$ and stellar mass $M_{\ast}$ are taken to be power law functions of halo mass:
\begin{align}
\label{escep_frec_eq}
    f_{\textrm{esc}} (M_{\rm h}) &= f_{\textrm{esc}, 10} \left(\frac{M_{\rm h}}{10^{10} M_{\odot}}\right)^{\alpha_{\textrm{esc}}}, \\
    M_{\ast} (M_{\rm h}) &= f_{\ast, 10}\ \left(\frac{M_{\rm h}}{10^{10} M_{\odot}}\right)^{\alpha_\ast} \left( \frac{\Omega_b}{\Omega_m}\right) \ M_{\rm h} .
    \label{eq:smhm}
\end{align}
Here, $f_{\textrm{esc},10}$ is the ionizing photon escape fraction normalized to the value in halos of mass $10^{10} M_{\odot}$, $f_{\ast,10}$ is the fraction of galactic gas in stars also normalized to the value in halos of mass $10^{10} M_{\odot}$, and $\alpha_{\textrm{esc}}$ and $\alpha_{\ast}$ are the corresponding power law indices. Both $f_{\rm esc}$ and $f_{\ast} \equiv f_{\ast,10} \left(\frac{M_{\rm h}}{10^{10} M_{\odot}}\right)^{\alpha_\ast}$ have a physical upper limit of $1$. This model also assumes that the star formation rate can be expressed on average as the stellar mass divided by some characteristic time scale:
\begin{equation}
    \dot{M}_{\ast} (M_{\rm h}, z) = \frac{M_\ast}{H^{-1}(z) t_{\ast}},
\end{equation}
where $H(z)$ is the Hubble parameter and $t_{\ast}$ is the characteristic time-scale for star formation (with this definition, its value varies from zero to unity).

This six-parameter galaxy model ($f_{\ast, 10}$, $\alpha_{\ast}$, $f_{\textrm{esc}, 10}$, $\alpha_{\rm esc}$, $M_{\textrm{turn}}$, $t_{\ast}$) is able to capture the average properties of the faint galaxies that dominate the ionizing photon budget, both from theoretical models and observations (e.g. \citealt{  Behroozi2019, Ishigaki2018, Ma2020, Park2020, Bouwens2022}).  Further details about the code and the parametrization can be found in \citet{2011MNRAS.411..955M, Park2019} and \citet{2020JOSS....5.2582M}.

For a given combination of astrophysical parameters, {\tt 21cmFAST} outputs 3D lightcones of the relevant cosmological fields.  We thus compute the patchy kSZ signal by ray-tracing through the ionization, density and LoS velocity lightcones, directly calculating the integral in equation (\ref{general_equation}), accounting also for the angular evolution of ${\bf \hat{u}}$ \citep[][]{Mesinger2012}.

\begin{figure*}
	\centering
    \includegraphics[width=\textwidth]{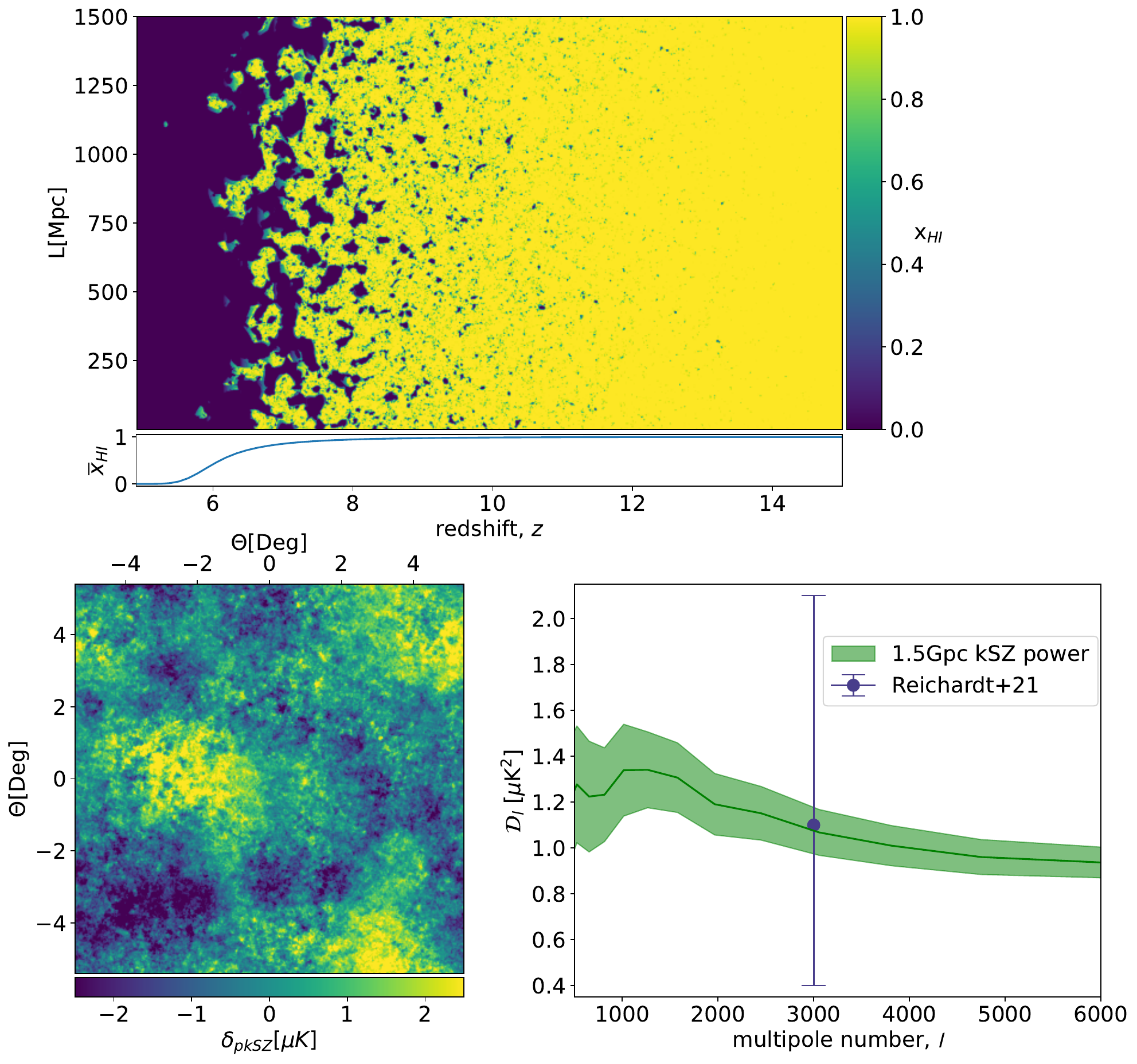}
    \caption{\textit{Upper panel:} 2D slice through the neutral hydrogen fraction lightcone together with its mean evolution on the bottom.  The lightcone slice is 1.5\,Gpc in height and 1.4\,Mpc thick. \textit{Lower left panel:} map of the patchy kinetic Sunayev-Zel'dovich signal, defined as being sourced by redshifts greater than five. \textit{Lower right panel:} Corresponding angular power spectrum of the patchy kSZ ({\it solid line}). The green shaded area highlights the 1$\sigma$ Poisson sample variance. Also shown is the recent measurement by \citet{Reichardt2021} at $l=3000$. This simulation used the following astrophysical parameters: $\log_{10}(f_{\ast, 10}) = -1.42$, $\alpha_{\ast} = 0.614$, $\log_{10}(f_{\textrm{esc}, 10}) = -1.78$, $\alpha_{\textrm{esc}} = 0.474$, $\log_{10}(M_{\textrm{turn}}) = 8.62$ and $t_{\ast} = 0.392$.}
    \label{kSZ plot}
\end{figure*}

In Fig. \ref{kSZ plot} we show an example of this procedure using a simulation that is 1.5 Gpc on a side. The astrophysical parameters of this simulation are taken from the posterior distribution of \citealt{Qin2021} (discussed further below), specifically:  [$\log_{10}(f_{\ast,10})$, $\alpha_\ast$, $\log_{10}(f_{\textrm{esc},10})$, $\alpha_{\textrm{esc}}$, $\log_{10}(M_{\textrm{turn}})$, $t_{\ast}$] = ($-1.42$, $0.614$, $-1.78$, $0.474$, $8.62$, $0.392$ ). The midpoint of EoR is at $z_{r} = 6.1$, while the neutral fraction drops to zero at $z_{\rm end} = 4.9$. The duration of the EoR, defined throughout as $\Delta_z \equiv z(\overline{x}_{\rm HI} =0.75)- z(\overline{x}_{\rm HI}=0.25)$, is $\Delta_z = 0.76$ and the CMB optical depth for the simulation is $\tau_e = 0.042$. In the top panel we show a 2D slice (with a thickness of $1.4$ Mpc) through the neutral fraction lightcone. In the bottom panels, we show the map of the patchy kSZ signal and the corresponding angular power spectrum.  While this model was chosen to have patchy kSZ power that agrees with the median estimate reported by \citealt{Reichardt2021}, complementary EoR and galaxy observations pull the posterior towards larger values of the $l=3000$ kSZ power,  as we quantify further below.

\subsection{Observations of the patchy kSZ}
\label{observing_section}

Observing the kSZ power spectrum is very challenging due to the presence of strong foregrounds as well as the primary CMB anisotropies.  Deep integration over multiple frequencies is essential in separating these different components of the power spectra.  
Over the past decade, ACT and SPT have published increasingly tighter upper limits on the cosmic kSZ signal \citep{Dunkley2013, Das2014, Reichardt2012, George2015}. Using SPT-SZ and SPTpol measurements at 95, 150 and 220 {GHz}, combined with a prior on the CIB-tSZ foregound from \citet{Crawford2014}, \citet{Reichardt2021} recently claimed a $3\sigma$ measurement of the total kSZ power: $\mathcal{D}_{3000}^{\rm kSZ} = 3.0 \pm 1.0 \ \mu\textrm{K}^2$ ($68\%$ C.I.).

 To isolate the patchy contribution to this total kSZ power, the authors subtracted an estimate of the $z<5.5$ homogeneous component 
based on the simulations of \citet{Shaw2012}: $\mathcal{D}_{3000}^{{\rm hkSZ}} = 1.65 \mu$K$^2$. The uncertainty around this value is bracketed 
by rescaling the best guess by a factor of $0.75$ and $1.25$. 
Doing so and using the bispectrum prior on tSZ, \citet{Reichardt2021} estimate the patchy kSZ power at $l=3000$ to be $\mathcal{D}_{3000}^{\textrm{pkSZ}} = 1.1 ^{+1.0}_{-0.7} \ \mu\textrm{K}^2$ ($68\%$ C.I.). 
Since our choice of lower bound in this work is $z=5.0$ instead of $z=5.5$, we add to the patchy kSZ estimate from \citet{Reichardt2021} the contribution of the homogeneous component over the redshift interval $5<z<5.5$. We estimate this be approximately $0.1 \mu$K$^2$ (e.g. fig. 6 in \citealt{Shaw2012}; fig. 5 in \citealt{Mesinger2012}).
Therefore we use the following observational constraint when performing inference in Section~\ref{sec:inference_set}: $\mathcal{D}_{3000}^{\textrm{pkSZ}} = 1.2 ^{+1.0}_{-0.7} \ \mu\textrm{K}^2$ ($68\%$ C.I.).



A more robust foreground model and a consistent analysis across scales can improve constraints, as demonstrated in \citet{Gorce2022} where the authors give an upper limit of $\mathcal{D}_{3000}^{\textrm{pkSZ}} < 1.58\,\mu\mathrm{K}^2$ (95\% C.L.) using the same data as \citet{Reichardt2021}.\footnote{As this project was started before the publication of \citet{Gorce2022}, here we use the original patchy kSZ estimate by \citep{Reichardt2021}.  The estimate in \citet{Gorce2022} would imply an even later reionization than shown here, also consistent with the newest analysis of the Lyman alpha forest spectra (Qin et al. in prep) as well as the forest dark fraction (\citealt{Jin2023ApJ...942...59J}; Campo et al. in prep). 
 We aim to revisit this in future work when more of these new constraints become public.}  Reducing the uncertainties on the total kSZ require deeper integration, lower noise levels, and more frequency channels to better characterize foregrounds and systematics, which future telescopes such as CMB-S4 and the Simons Observatory (e.g. \citealt{Abitbol2017, Ade2019}) are expected to achieve. Furthermore, robustly isolating the patchy component of the total kSZ signal requires exhaustively sampling models of galaxy clusters in order to better characterize the post-reionization (homogeneous) component.  Motivated by upcoming data and improved analysis, we also perform a forecast run from a mock measurement with error bars corresponding to the uncertainty expected from future experiments. This is presented in \autoref{sec:forecast}.

\section{Complementary EoR and galaxy observations}
\label{sec:complimentary_observations}

We now have several, independent observational probes of the EoR which can help constrain astrophysical parameters \citep[e.g.][]{Choudhury2006, Greig2015, Gorce2018, Park2019}.
Here we follow \citet{Qin2021}, who used the same galaxy parametrization as we do, and use the following observational data:
\begin{enumerate}

\item {\it Lyman $\alpha$ forest opacity distributions} -- the $5.4 \leq z \leq 6.0$ probability density functions (PDFs) of the forest effective optical depth, $\tau_{\rm eff} \equiv - \ln \langle f \rangle_{\rm 50\,Mpc}$, computed from the mean normalized flux, $f$, of the QSO sample in \citet{Bosman2018}.  \citet{Qin2021} showed that this data require reionization to end late, $z \leq 5.6$ (see also \citealt{Choudhury2021b}). 

\item {\it Dark fraction in the Ly$\alpha$ and Ly$\beta$ forests} -- the fraction of QSO spectral pixels that are dark (zero transmission) in both Lyman alpha and Lyman beta from the sample in \citet{McGreer2015}.  This so-called dark fraction provides a model-independent upper limit on the neutral hydrogen fraction, with the  value at $z\sim5.9$ corresponding to $\bar{x}_{\rm HI} < 0.06 + 0.05$ (1$\sigma$). 
This dataset favors earlier reionization models.\\

\item {\it High-redshift galaxy UV luminosity functions (UV LFs)}  -- the 1500 \AA\ restframe UV LFs at $z=6-10$, estimated by \citet{Bouwens2015, Bouwens2016, Oesch18}. To constrain our models, we assume a conversion factor between the star formation rate (SFR) and UV luminosity, $\dot{M}_{\ast} = \mathcal{K}_{\rm UV} L_{\rm UV}$, and take $\mathcal{K}_{\rm UV} = 1.15 \cdot 10^{-28}\, \textrm{M}_{\odot} \textrm{yr}^{-1} \textrm{erg}^{-1} \ \textrm{s} \ \textrm{Hz}$, following \citet{Sun2016}\footnote{This value was obtained assuming a stellar metallicity of $Z_{\ast} = 10^{-0.15z}Z_{\odot}$ and a Salpeter initial mass function (see also \citet{Madau2014}). 
}. UV luminosities are then related to magnitudes using the AB magnitude relation \citep{Oke1983}: $\log_{10}\left(\frac{L_{\rm UV}}{\textrm{erg} \textrm{s}^{-1}\textrm{Hz}^-1}\right) = 0.4 \times (51.63 - M_{1500})$. 
UV LFs are very useful in anchoring our SFR relations (i.e. the ratio $f_\ast/t_\ast$), using the more massive reionization-era galaxies bright enough to be observed directly with the {\it Hubble} (and eventually {\it JWST}) telescope.\\

\item {\it The CMB optical depth} -- the Thomson scattering optical depth of CMB photons as computed by \citet{Planck2020}, $\tau_e = 0.0561 \pm 0.071$ ($1\sigma$).  Although it is more accurate to directly forward model the CMB EE power spectra, \citet{Qin2020} show that computing the likelihood from $\tau_e$ (a compressed summary statistic of the CMB power spectra) does not notably impact the resulting posterior for the astrophysical model used here. 
\end{enumerate}

These four complementary datasets are used in all of our inferences, each contributing a factor in the final likelihood. We write out explicitly all likelihood terms in Appendix~\ref{appendixB}. For further details, we refer the interested reader to \citet{Qin2021}.

\section{WHAT DO WE LEARN FROM THE PATCHY KSZ SIGNAL?}
\label{results}

We now explore what astrophysical constraints can be obtained from reionization observations, including the recent kSZ measurement \citep{Reichardt2021}.  We first discuss our  Bayesian sampler and the set up of our forward models, before showing results using current and future kSZ measurements.

\subsection{Inference set-up}
\label{sec:inference_set}


To perform Bayesian inference, we use {\tt 21cmMC}\footnote{Available at \url{https://github.com/21cmfast/21CMMC}.} \citep{Greig2015, Greig2018}, a public Monte Carlo sampler of {\tt 21cmFAST}.  For each set of model parameters (see section \ref{estimating_section}), {\tt 21cmMC} computes a 3D lightcone realization of cosmological fields, comparing the model to the observations (see sections \ref{observing_section} and \ref{sec:complimentary_observations}).  Here we use the MultiNest \citep{Feroz2009, Qin2021b} sampler, which is fully implemented in {\tt 21cmMC} and scales well to high-dimensional inference \citep[e.g.][]{HERA22}.  We use $1000$ live points, an evidence tolerance of $0.5$ and a sampling efficiency of $0.8$. We checked for convergence by launching a run with $2000$ live points and found no significant difference in the inferred posterior distributions. Our fiducial posterior converges after $\sim 45$k samples, taking $\sim 260$k core hours.

Unfortunately, due to computational limitations, we cannot use ultra-large simulations (e.g. Fig.~\ref{kSZ plot}) when forward modeling.  Instead we use smaller boxes, calibrating their output to account for the missing large-scale modes in the kSZ signal (see also \citealt{Iliev2007, Shaw2012, Park2013, Alvarez2016}).  Specifically, we use simulations of ($500$\,Mpc)$^3$ on a $256^3$ grid. When constructing the lightcones, we rotate the coeval boxes to minimize duplication of structures due to periodic boundary conditions (e.g. \citealt{Mesinger2012}). We account for the missing large-scale power by sampling several realizations (different cosmic seeds) of $500$\,Mpc boxes, and comparing their power spectra to those from $1.5$\,Gpc boxes, constructed using the same astrophysical parameters. We compute the mean ratio of the missing power, $f_{0.5\textrm{Gpc}}^{1.5\textrm{Gpc}} \equiv \mathcal{D}_{3000}^{\textrm{pkSZ}, 1.5\textrm{Gpc}} / \mathcal{D}_{3000}^{\textrm{pkSZ}, 500\textrm{Mpc}}$, adjusting our forward models by this factor and including the corresponding variance in the denominator of the likelihood. We obtain $f_{0.5\textrm{Gpc}}^{1.5\textrm{Gpc}}$ = $1.27 \pm 0.19$. Further details on this calibration procedure can be found in Appendix~\ref{appendixA}. 

\subsection{Inference results using the recent SPT measurement}
\label{sec:inference}

We compute two posteriors:

\begin{figure*}
	\includegraphics[width=\textwidth]{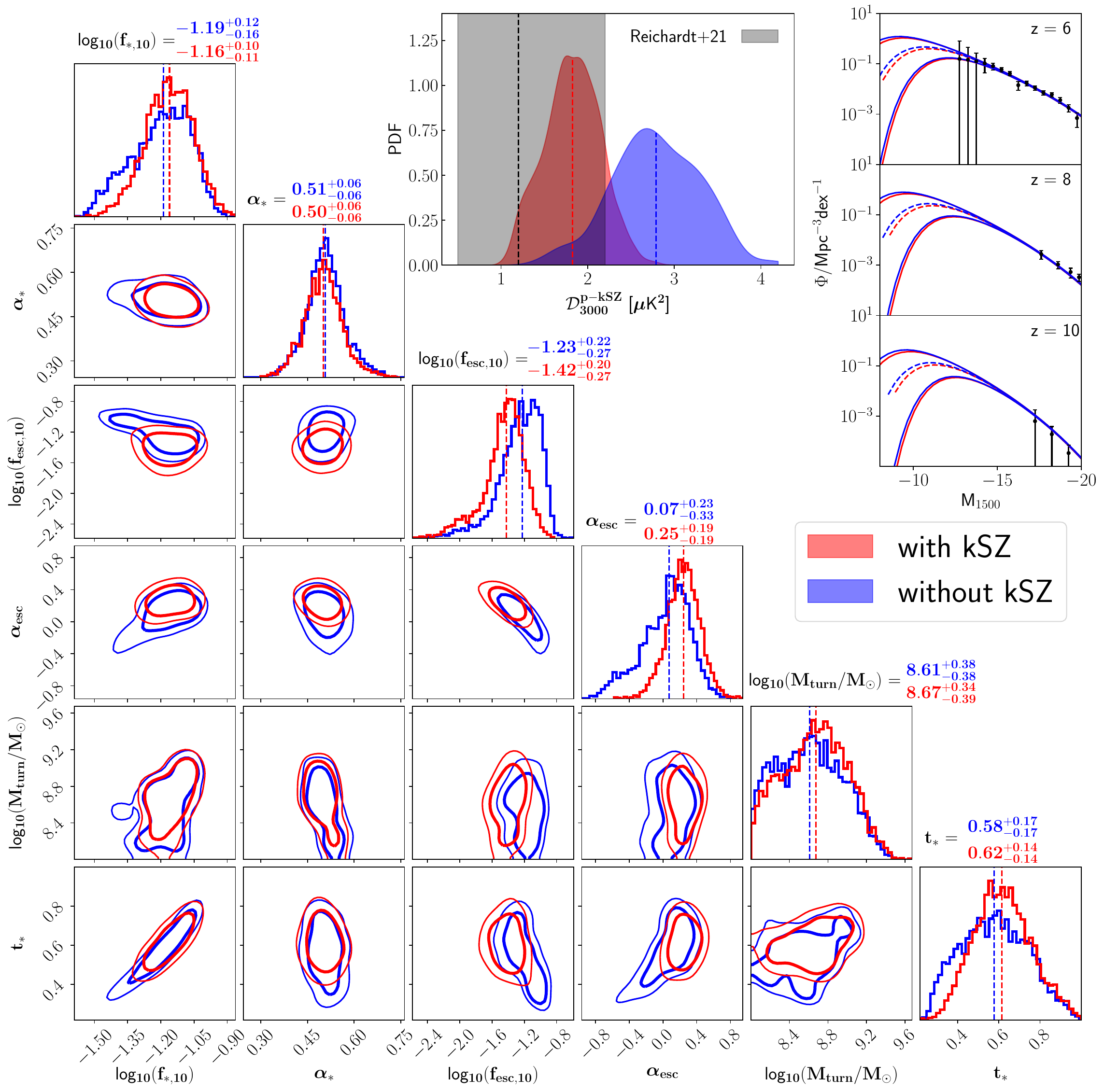}
    \caption{Marginalized posteriors of {\it without kSZ} ({\it blue}) and {\it with kSZ} ({\it red}).  As discussed in the text, {\it without kSZ} is constrained using large-scale Ly$\alpha$ forest opacity PDFs, the forest dark fraction, UV LFs, and the CMB optical depth, while {\it with kSZ} additionally includes the recent measurement of the patchy kSZ power at $l=3000$.
    The 1D and 2D posterior distributions of the model parameters are shown in a corner plot on the left, with thin and thick lines representing 95\% and 68\% credible intervals (C.I.) respectively. The marginalized median values (shown also as dashed lines) with the 68\% central C.I. are given over the 1D distribution functions for the two runs. In the upper middle panel we show the PDFs of the patchy kSZ signal power spectrum at $l=3000$, together with the \citet{Reichardt2021} observational estimate in grey. 
    Also shown are the median and [14, 86]\%~C.I. (dashed and solid lines respectively) of the inferred UV luminosity functions at $z$ = 6, 8 and 10. Black points with error bars are UV LF observations used for the inference from \citet{Bouwens2015, Bouwens2016} and \citet{Oesch18}.}
    \label{corner_plot}
\end{figure*}

\begin{figure*}
    \includegraphics[width=\textwidth]{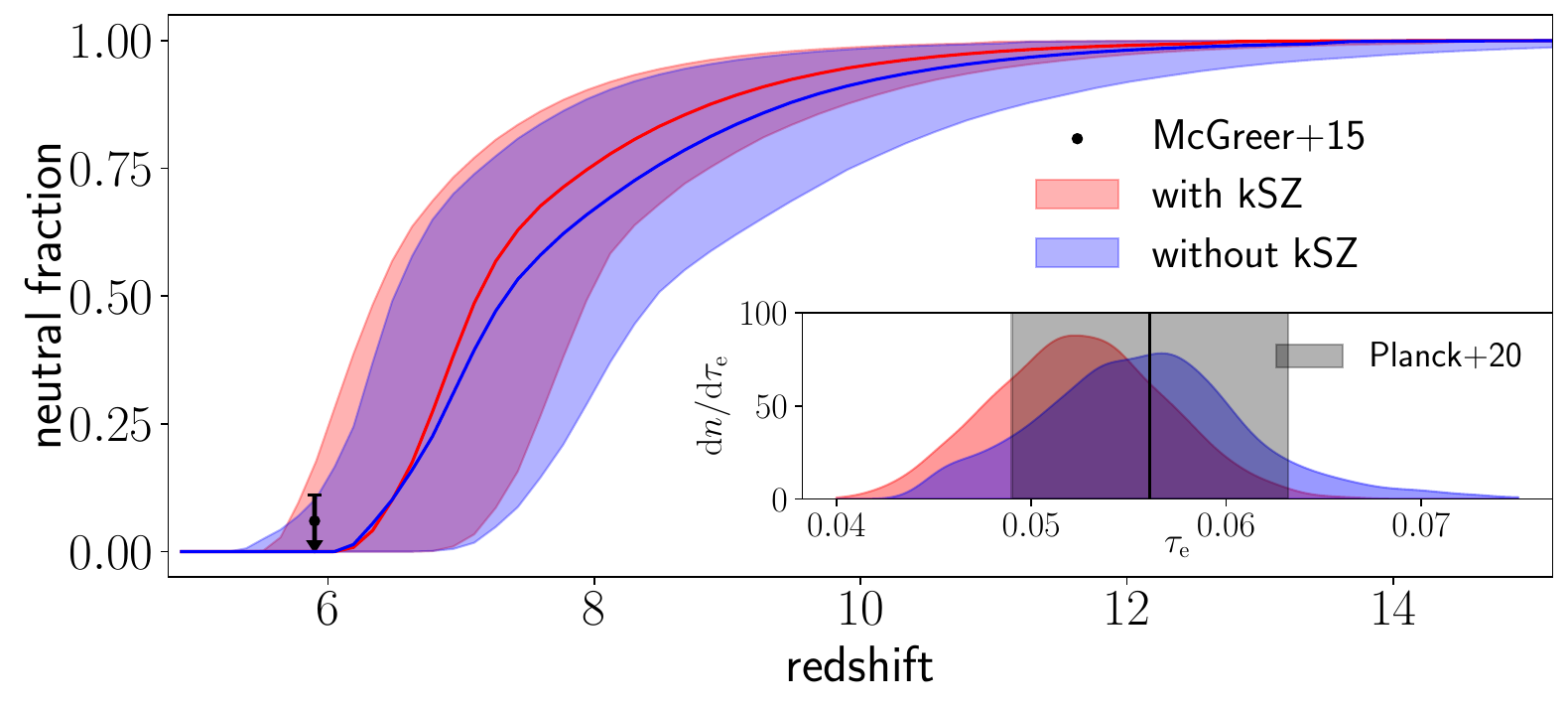}
    \caption{Median EoR histories with 95\% C.I. for the {\it without kSZ} ({\it blue}) and {\it with kSZ} ({\it red}) posteriors. Also shown is the upper limit from \citet{McGreer2015} at $z \sim 5.9$ that is used in the likelihood (Section \ref{sec:complimentary_observations}) for both posteriors. The insert shows probability density distributions of $\tau_{\rm e}$ for the \textit{with kSZ} and \textit{without kSZ} posteriors. The vertical black line and gray shaded region correspond to the \citet{Planck2020} measurement of $\tau_{\rm e}$, also used in the likelihood for inferences. Including kSZ data increasing the preference for a later and more rapid reionization.}
    \label{fig:xh_evol}
\end{figure*}

\begin{itemize}

\item {\it without kSZ} -- this corresponds to the posterior based on the observational data (i)--(iv) from the previous section, i.e. large-scale Ly$\alpha$ forest opacity PDFs, the forest dark fraction, UV LFs, and the CMB optical depth.\footnote{Even though we used the same parametrization and observational data as \cite{Qin2021}, our {\it without kSZ} posterior distribution is slightly different.  This is because here we use the ionizing photon conservation correction from \citet{Park2022}, which results in roughly a shift of 0.2 in the recovered $\alpha_{\rm esc}$ (as also shown in \citealt{Park2022}).  When computing the Lyman alpha forest we use a harder UV background (with energy index $\beta_{\rm uv}=-2$ instead of $-5$) and a higher post-ionization front temperature ($T_{\rm re} = 2.0\times10^4$K instead of $1.0\times10^4$K), motivated by recent estimates from hydrodynamic simulations (e.g. \citealt{Daloisio2019}).  The harder UV background shifts the end of reionization to slightly earlier times, compared with \cite{Qin2021}.}

\item {\it with kSZ} -- this is the same as {\it without kSZ}, but including an additional factor in the likelihood, $\mathcal{L}_{\rm kSZ}$ (see Appendix~\ref{appendixA} for details) corresponding to the patchy kSZ measurement by \citet{Reichardt2021}, adjusted for the slightly different lower redshift bound as discussed above: $\mathcal{D}_{3000}^{\textrm{pkSZ}} = 1.2 ^{+1.0}_{-0.7} \ \mu\textrm{K}^2$ ($68\%$ C.I.).
\end{itemize}

Comparing the {\it without kSZ} and {\it with kSZ} posteriors, we quantify the additional constraining power provided by the patchy kSZ.  We  begin by showing the constrains on the fundamental galaxy parameters, before discussing the corresponding derived quantities such as the EoR history and the halo-galaxy connection.

\subsubsection{Galaxy parameters and EoR history}
\label{parameter_inference}

In the bottom left of Fig.~\ref{corner_plot}, we show the resulting two- and one-dimensional posteriors {\it without kSZ} (blue) and {\it with kSZ} (red).  We also show the model posteriors together with two of the observational data used in the likelihood: the $l=3000$ patchy kSZ power ({\it top center}), and the UV LFs at $z=6$, 8, 10 ({\it top right}).

From the $\mathcal{D}_{3000}^{\textrm{pkSZ}}$ PDFs shown in the top center panel, we see that the recent measurement by \citet{Reichardt2021} is in mild tension with the {\it without kSZ} posterior:  the kSZ data favor the low amplitude tail of the  {\it without kSZ} posterior, corresponding to late reionization models. 
%
Indeed, by including the kSZ measurement, the distribution  is shifted in favor of smaller kSZ power.  Most of the {\it with kSZ} posterior is still above the mean estimate of the kSZ power by \citet{Reichardt2021}, though perfectly consistent given the large observational uncertainty.

The biggest difference between the two galaxy parameter posteriors is in the recovered ionizing escape fraction, parametrized in our model with $f_{\textrm{esc},10}$ and $\alpha_{\textrm{esc}}$ (see Eq.~\ref{escep_frec_eq}). The SPT measurement favors slightly lower values of $f_{\textrm{esc, 10}}$ and higher values of $\alpha_{\rm esc}$. As a result, the inferred ionizing efficiency slightly increases in more massive, late-appearing galaxies (discussed further in the following section), so that the EoR occurs later and more rapidly, as can be seen from the EoR histories shown in Fig. \ref{fig:xh_evol}. Including the relatively low patchy kSZ amplitude claimed by \citet{Reichardt2021} disfavors the more extended EoR histories present in the {\it without kSZ} posterior. While the end of the EoR remains fairly unchanged, constrained by Lyman alpha forest observations (e.g. \citealt{Qin2021, Choudhury2021b}), the middle and early stages are shifted to later times with the addition of kSZ data.  This translates into lower CMB optical depths as seen in the inset of Fig. \ref{fig:xh_evol}: $\tau_e = 0.052_{-0.008}^{+0.009}$ for the \textit{with kSZ} posterior compared to $\tau_e = 0.055_{-0.009}^{+0.012}$ for the \textit{without kSZ} ome. 
The EoR histories implied by the \textit{with kSZ} posterior are consistent with the Lyman $\alpha$ forest data, but in slight ($\lesssim 1 \sigma$) tension with the CMB optical depth $\tau_e$ as well as the QSO dark pixel fraction. We note that an updated estimate of the QSO dark pixel fraction using more recent, much larger QSO samples from \citet{Dodorico+23} results in weaker upper limits on the neutral fraction at $z\sim6$, making them perfectly consistent with later EoR models (Campo et al. in prep).  This would leave the CMB $\tau_e$ as the only dataset preferring a slightly earlier EoR. Such a mild tension between the two CMB datasets could come from calibration or analysis inconsistencies between large- and small-scale data, that is between the SPT and Planck data \citep[e.g.][]{Gorce2022}.


On the other hand, constraints on parameters governing the star formation rates and stellar-to-halo mass relations (i.e. $f_{\ast, 10}$, $\alpha_\ast$, $t_\ast$, $M_{\rm turn}$) are fairly unchanged when including kSZ data. As already shown in \citealt{Park2019, Qin2021}, observed high-$z$ UV LFs constrain the stellar-to-halo mass relation ($f_{\ast, 10}$, $\alpha_\ast$) and place an upper limit on a faint end turnover ($M_{\rm turn}$). Therefore, these parameters have only limited freedom to impact the timing of reionization while still being consistent with the UV LFs data.  

\subsubsection{Scaling relations of EoR galaxies}
\label{sec:scaling}

\begin{figure*}

    \includegraphics[width=0.9\textwidth]{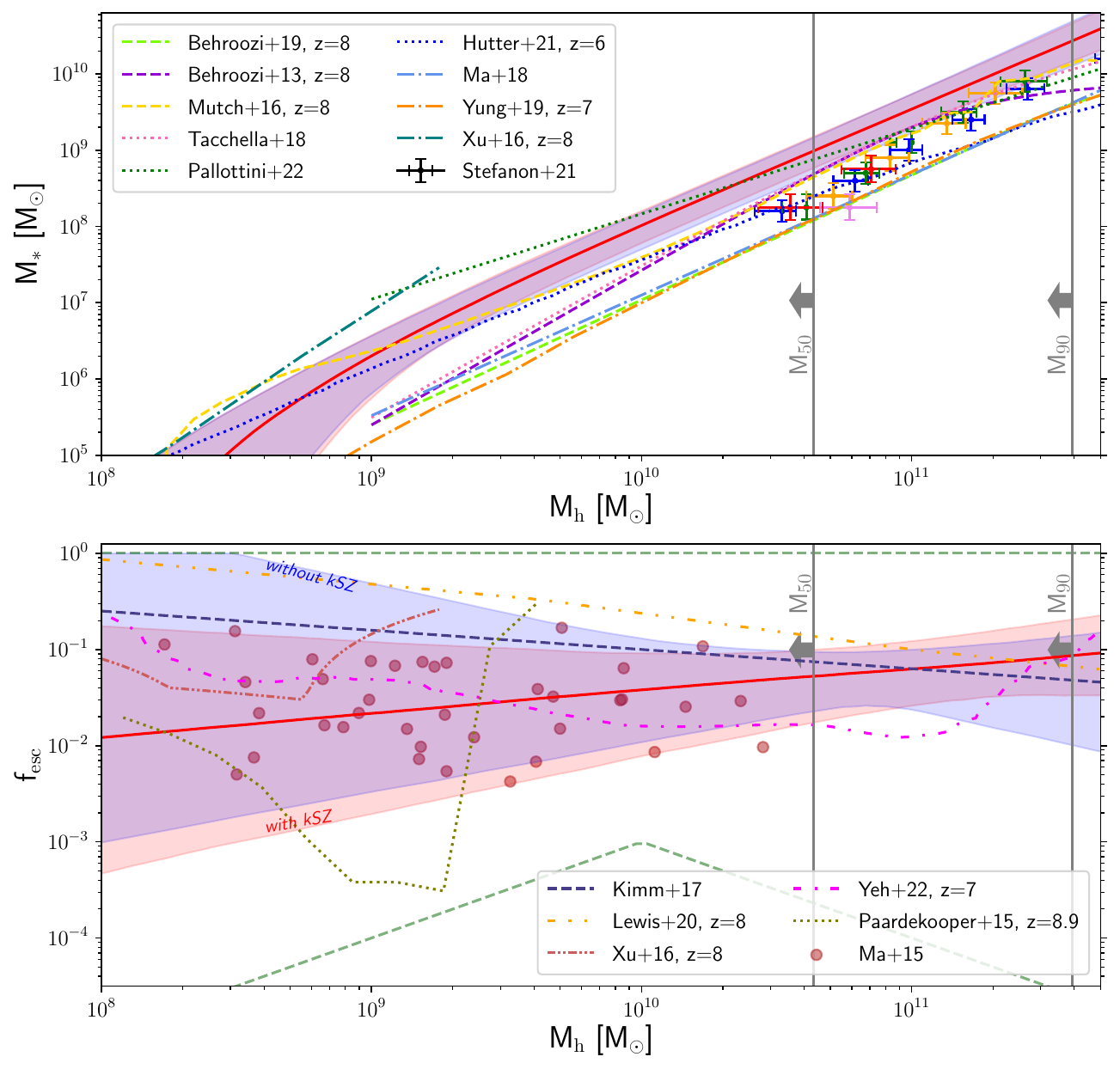}
    \caption{
    The dependence of average galaxy properties with halo mass. Shaded regions represent 95\% C.I. of  {\it without kSZ} ({\it blue}) and {\it with kSZ} ({\it red}).  Vertical lines demarcate the posterior-averaged mean of $M_{50}$ ($M_{90}$), defined as halo mass upper limit below which galaxies source 50\% (90\%) of the ionizing emissivity at $z=7$.
    \textit{Upper panel:} stellar to halo mass relation (SHMR). For illustrative purposes, we also show a selection of independent results from empirical and semi-analytic models \citep{Behroozi2013, Mutch2016, Tacchella2018, Behroozi2019, Yung2019, Hutter2021}, as well as from cosmological, hydro simulations \citep{Xu2016, Ma2018, Pallottini2022}. For \citet{Tacchella2018} and \citet{Ma2018}, we use the redshift independent fits from their Eq.~$12$ and Eq.~$1$, respectively. The \citet{Xu2016} result is taken for their Void region at z$=8$, though the SHMR is similar for other environments and redshifts (see their fig.~$16$). The \citet{Pallottini2022} curve is a linear fit to their data points. Colored points with error bars correspond to abundance matching estimates assuming a constant duty cycle \citep{Stefanon2021}, where blue/green/orange/red/pink points are for z$=6$/$7$/$8$/$9$/$10$ .
    \textit{Lower panel:} the ionizing escape fraction. The green dashed lines demarcate the prior range.  Again for illustrative purposes, we show estimates from some cosmological, hydrodynamic simulations \citep{Paard2015, Ma2015, Xu2016, Kimm2017, Lewis2020, Yeh2022}. For \citet{Kimm2017} we show their "fiducial" model.  The \citet{Ma2015} points represents time-averaged escape fractions obtained using their SHMR relation (see their figures~3 and 9.)}
    \label{fig:shmh_esc}
\end{figure*}

To gain further insight into the implications of our results, we show the corresponding galaxy scaling relations in Fig.~\ref{fig:shmh_esc}. In the top panel, we plot the inferred stellar-to-halo mass relation (SHMR), defined as the average stellar mass inside a halo of mass $M_{\rm h}$ (including the $f_{\rm duty}$ occupation fraction term from Eq.~\ref{eq:fduty}).  The redshift-independent median relation from the {\it with kSZ} posterior is denoted with the red solid line, corresponding to:
\begin{equation}
\frac{\bar{M}_\ast}{M_h} = 0.011_{-0.002}^{+0.003} \left( \frac{M_{\rm h}}{10^{10} \textrm{M}_{\odot}} \right)^{0.50_{-0.06}^{+0.06}} \quad  (68\% \textrm{C.I.}) ,
\end{equation}
while the 95\% C.I. is shown with the purple shading.\footnote{We note that the inferred median scaling of the SHMR $\propto M_h^{0.5}$, is close to the value expected by simply assuming SNe feedback scales with the gravitational potential of the host halo, SHMR $\propto M_h^{0.67}$ \citep[e.g.][]{Wyithe2013}. As discussed further below, any mass dependence of $\mathcal{K}_{\rm UV}$ (here assumed to be constant) would also impact the inferred scaling.}.  We also show the corresponding 95\% C.I. of the {\it without kSZ} posterior in blue.  The two posteriors overlap in this space, again illustrating that the SHMR for our model is determined by the UV LFs, and is unaffected by kSZ data.
  We also include some other estimates from the literature, which show sizable scatter for the high-redshift, small-mass
regime that is relevant for the EoR.   Our inferred relation is roughly consistent with current estimates, given their large scatter.  It is unsurprising that, despite the fairly large scatter, the slopes of the SHMRs shown in this panel are roughly similar.  This is because in most cases the observed UV LFs are used either directly or indirectly to calibrate the models.  The slope of the UV LFs combined with the slope of the HMF, both power-laws in this range, sets the slope of the SHMR, with the normalization being more sensitive to the star formation -- $L_{1500}$ conversion\footnote{
We caution that our SHMRs are likely overconstrained, because we do not include any uncertainty in the SFR -- $L_{1500}$ conversion (i.e. we fix $\mathcal{K}_{\rm UV}$ from Sec.~\ref{sec:complimentary_observations} to be a constant).  This conversion depends on the IMF and the duration of recent star formation, with different assumptions changing $\mathcal{K}_{\rm UV}$ by factors of $\sim 2$ (e.g. \citealt{Wilkins2019, Stanway2020}).}.

Similarly, in the bottom panel of Fig. \ref{fig:shmh_esc}, we show the ionizing escape fraction to halo mass relation.  Our redshift-independent median relation for \textit{with kSZ} is denoted with a solid red line:
\begin{equation}
f_{\rm esc} = 0.038_{-0.017}^{+0.021} \left( \frac{M_{\rm h}}{10^{10} \textrm{M}_{\odot}}\right)^{0.25_{-0.19}^{+0.19}} \quad (68\% \textrm{C.I.}; \textit{with kSZ}),
\end{equation}
Also shown is the result for \textit{without kSZ} in blue:
\begin{equation}
    f_{\rm esc} = 0.060_{-0.028}^{+0.038} \left( \frac{M_{\rm h}}{10^{10}\textrm{M}_{\odot}}\right)^{0.07_{-0.33}^{+0.23}} \quad (68\% \textrm{C.I.}; \textit{without kSZ}).
\end{equation}
As in the panel above, the corresponding shaded areas demarcate the 95\% C.I.  The green dashed lines denote the range of our prior in this space, uniform over $\log_{10} f_{\rm esc, 10} \in [-3, 0]$, $\alpha_{\rm esc} \in [-1, 0.5]$; the fact that our posterior is tighter than the prior illustrates the constraining power of current observations and that our results are not sensitive to our choice of prior. 

Again for illustrative purposes, we show some theoretical estimates from the literature.  Compared to the SHMR in the top panel, there is far less consensus on the ionizing escape fraction.  This is because the relevant small scales are impossible to resolve in cosmological simulations; therefore results are sensitive to the resolution/sub-grid prescriptions.  Indeed, some simulations suggest an increasing trend with halo mass while others suggest a decreasing trend.

In contrast, Bayesian inference allows the observations to inform us about the (mean) $f_{\rm esc}(M_{\rm h})$ relation.  By comparing the blue and red shaded regions we see that the addition of kSZ data favors a slight increase in the mean escape fraction towards more massive halos.  While the uncertainties are still large at the small mass end, the ionizing escape fraction for galaxies hosted by $\sim 10^{10}$ -- $10^{11}$ $M_\odot$ halos is reasonably well constrained to be a few percent.   Interestingly, strong evolution with halo mass is disfavored.

In Fig. \ref{fig:shmh_esc} we also demarcate the posterior-averaged mean of $M_{50}$ ($M_{90}$), defined as halo mass upper limit below which galaxies source 50\% (90\%) of the ionizing emissivity at $z=7$, i.e. $M_{50}$ is calculated by solving the equation: $\int_{0}^{{\rm M}_{50}} \rm{d} \textrm{M}_{\rm h} \ {\rm d n}_{{\rm ion}}/{\rm d M_{\rm h}}  = 1/2 \int_{0}^{\infty}  \rm{d} \textrm{M}_{\rm h} \ {\rm d n}_{{\rm ion}}{\rm d M_{\rm h}}$.  We see that over half of the ionizing photons are sourced by galaxies that are below current detection limits.

\subsection{Forecast assuming future kSZ measurments}
\label{sec:forecast}

The improved precision of future experiments, as well as their larger sky coverage, will allow for lower noise levels and decreased sample variance. With CMB-S4, we expect the errors on the measurement of the amplitude of the CMB temperature power spectrum at $l = 3000$ to decrease by a factor of $5$ to $10$, depending on the bandpower \citep{Abazajian2016}. Improved foreground modelling should also help reduce the uncertainty on the kSZ amplitude by roughly $30 \%$ \citep{Gorce2022}. On the theoretical side, suites of simulations could better characterize the contribution of the homogeneous kSZ to the total power.

To quantify the corresponding improvement in parameter constraints, we repeat the {\it with kSZ} inference in Sec.~\ref{parameter_inference}, but using a mock future kSZ measurement instead of \citet{Reichardt2021}. 
We assume $\mathcal{D}_{3000}^\mathrm{pkSZ} = 2.0 \pm 0.10\,\mu \textrm{K}^2$. The mean value corresponds to the maximum a posteriori (MAP) model from the {\it with kSZ} posterior in the previous section, while the choice of uncertainty is (very roughly) motivated by the arguments above.

\begin{figure}
	\includegraphics[width=\columnwidth]{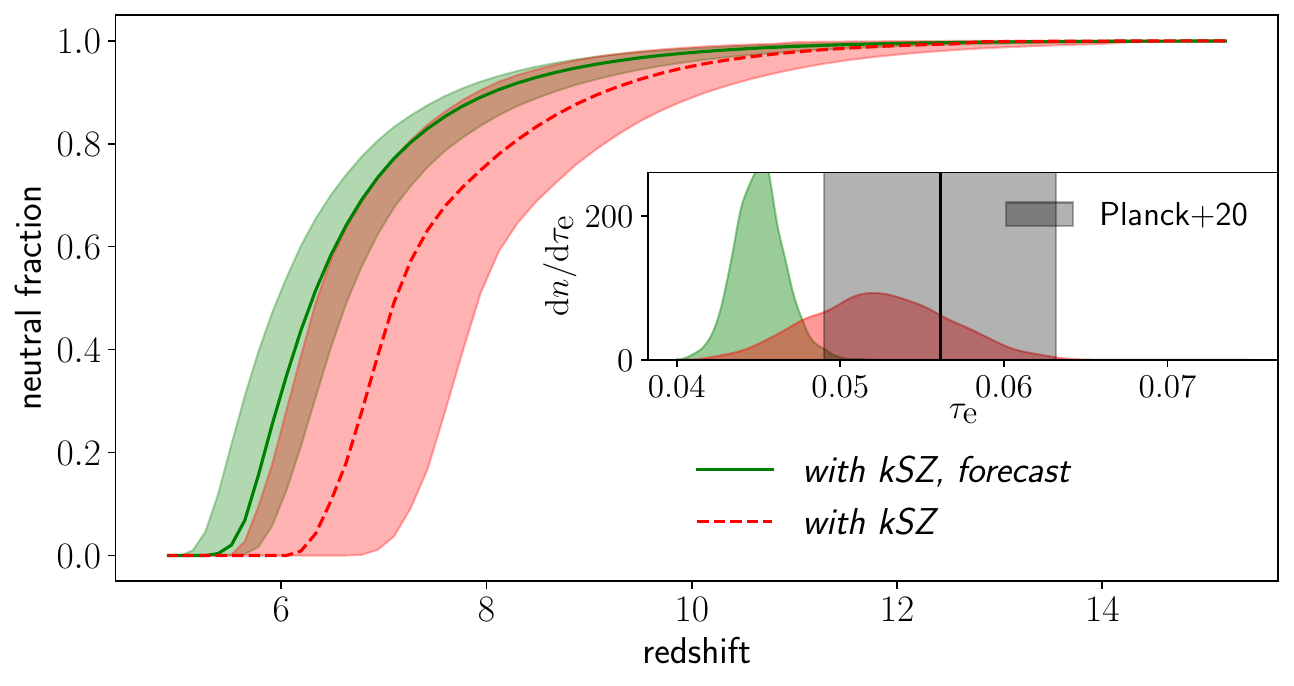}
    \caption{Median EoR histories with 95\% C.I. for the {\it with kSZ} ({\it red}) posterior and new posterior with the forecast value of patchy kSZ amplitude: $\mathcal{D}_{3000}^\mathrm{pkSZ} = 2.0 \pm 0.10\,\mu \textrm{K}^2$ ({\it green}). In the inset we also show the corresponding PDFs of $\tau_{\rm e}$.}
    \label{fig:forecast_xh}
\end{figure}

In Figure \ref{fig:forecast_xh} we show constraints on the EoR history using the mock kSZ observation ({\it green shaded region}), together with the current constraints ({\it red shaded region}).
We see that if the kSZ error bars could be reduced by a factor of $\sim 10$, it would result in a dramatic improvement on the recovered EoR history, with the
midpoint of reionization being constrained to an r.m.s. uncertainty of $\sigma_{z_r} = 0.16$, compared to $0.4$ for the current \textit{with kSZ} posterior in red. A similar improvement is also obtained for the duration of EoR: $\Delta_z \equiv z(\overline{x}_{\rm H} =0.75)- z(\overline{x}_{\rm H}=0.25)$. Using the mock kSZ observation we recover $\Delta_z = 1.09_{-0.09}^{+0.12}$, which compared to the current \textit{with kSZ} constraints of $\Delta_z = 1.16_{-0.19}^{+0.24}$, reduces the uncertainty by a factor of $\sim 2$.

\begin{figure*}
	\includegraphics[width=\textwidth]{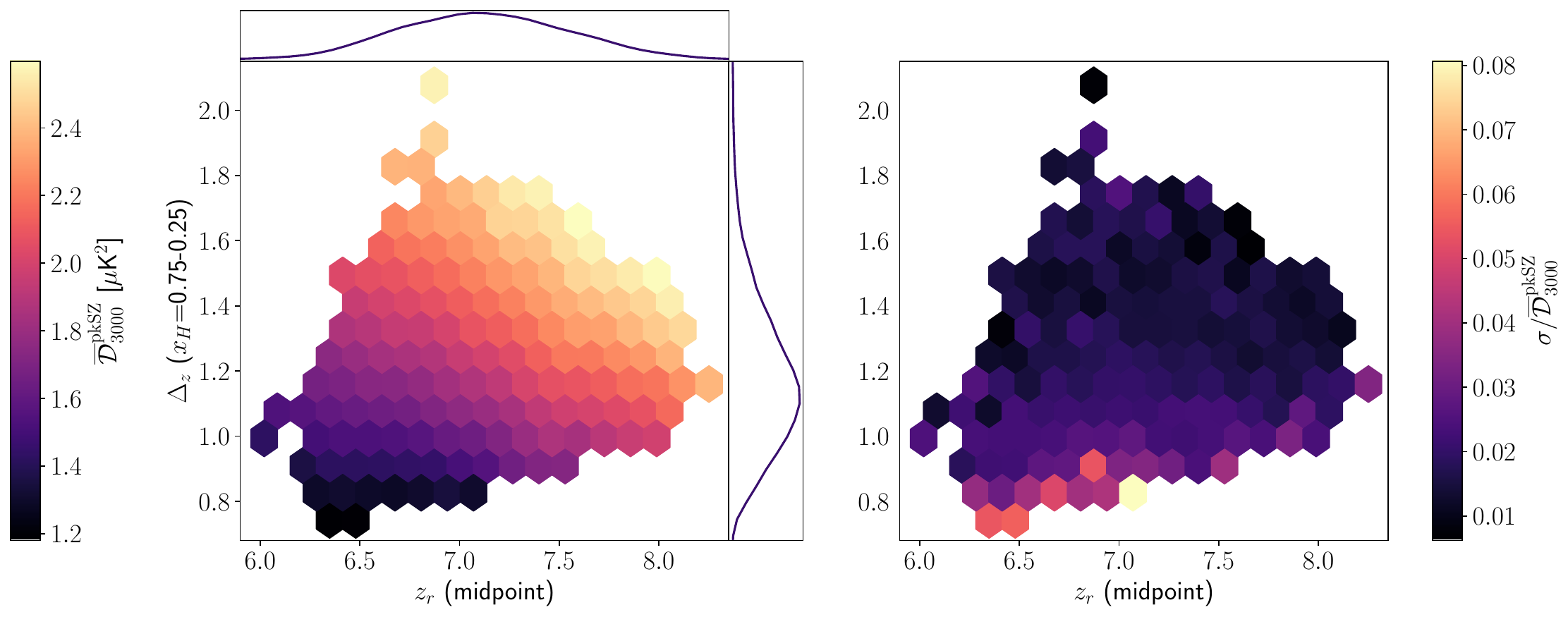}
    \caption{\textit{Left panel:} Mean value of the patchy kSZ power spectrum amplitude at $l=3000$ binned as a function of $z_r$ (midpoint of reionization) and $\Delta_z$ (duration of reionization; $\Delta_z \equiv z(\overline{x}_{\rm H} =0.75)- z(\overline{x}_{\rm H}=0.25$)). The samples are taken from the \textit{with kSZ} run. Plotted are the bins for which the scatter in the bin is larger than the uncertainty of the mean. 1D posterior distribution of $\Delta_z$ and $z_r$ are shown on the sides. \textit{Right panel:} Standard deviation of the patchy kSZ power spectrum at $l=3000$ within each bin divided by the mean for the same bin.}
    \label{mean_sigma}
\end{figure*}

It is interesting to note that the change in the recovered history is primarily in delaying reionization; the duration decreases only marginally.
Because galaxies sit inside halos, the duration of reionization cannot be arbitrarily short; it will be limited by the growth of the HMF. The most rapid EoR models are those dominated by the rare, bright galaxies hosted by massive halos in the exponential tail of the HMF. Their fractional abundance increases more rapidly compared to that of the more common, smaller halos.  However, the observed UV LFs set a lower limit on $\Delta_z$ because we actually see galaxies down to $M_{1500} \sim -13$, and the rare bright galaxies cannot have $f_{\rm esc} > 1$.  Since we cannot physically decrease $\Delta_z$ to values below unity, the only physically plausible way of decreasing the kSZ amplitude to agree with the mock observation is to lower the redshift of reionization \footnote{This picture can be changed somewhat if additional ionizing sources are present, such as AGN. A rapid increase in the number density of AGN could make reionization end somewhat more abruptly. However recent estimates imply AGN cannot contribute more than $\sim10$\% to the EoR (e.g. \citealt{Qin2017, Harikane2023}), so we expect their eventual impact to be modest. We also note that our escape fraction parametrization effectively captures EoR histories in which AGN contribute significantly through high values of the $\alpha_{\rm esc}$ parameter. In future work we will further increase the flexibility of our model, allowing for an explicit contribution from AGN.}.  This is seen in the figure, and it causes the CMB $\tau_e$ PDF to pile up in the lower $\sim 30$\% C.I. inferred from {\it Planck} alone.   Interestingly this later EoR history is in very good agreement with the latest, independent estimates coming from the Lyman alpha forest opacity fluctuations which imply that EoR finishes at $z \approx 5.3$ (Qin et al., in prep).

\section{Do we need self consistent forward models of the kSZ?}
\label{self:self-consistent}

When interpreting kSZ observations, it is common to vary the amplitude of the patchy kSZ power but with a fixed power spectrum shape (e.g. \citealt{Zahn2012, Battaglia2013, Reichardt2021}). Generally the power at $l=3000$, $\mathcal{D}^{\rm pkSZ}_{3000}$, is related to empirical parameters characterizing the EoR history, such as its midpoint and duration (with several definitions found in the literature).  This is in contrast to our approach in which the patchy kSZ power spectra are self-consistently forward-modeled directly from galaxy properties.

Using only empirical parameters for the EoR history has two important drawbacks: (i) for a given EoR history, the patchy kSZ power can also vary due to the EoR {\it morphology} \citep{Mesinger2012, Battaglia2013, Gorce2020, Paul2021, Choudhury2021, Chen2022}; and (ii) it is more difficult to physically-motivate priors for {\it derived} EoR history parameters, than it is for the fundamental galaxy parameters \citep[e.g.][]{Qin2020}.  The choice of priors is especially important when the likelihood is not overly constraining (e.g. \citealt{Trotta2017, Efstathiou21}).\footnote{In the previous section, we showed that our likelihood was indeed quite constraining, and therefore our posterior was not sensitive to our choice of priors.   This is because we use several complementary EoR and galaxy observations to construct the likelihood.  However, when only using the kSZ observation and ignoring for example the UV LFs, the likelihood is not overly constraining and the posterior can strongly depend on the choice of priors over the EoR history parameters \citep[e.g.][]{Greig2017, Park2019, HERA22}.}
Here we briefly explore the impact of (i). We sample our {\it with kSZ} posterior from Section \ref{sec:inference}, computing for each sample the midpoint of the EoR, $z_r$, and its duration, $\Delta_z$.  In Fig.~\ref{mean_sigma} we plot the mean ({\it left panel}) $\bar{\mathcal{D}}^{\rm pkSZ}_{3000}$ and normalized r.m.s. ({\it right panel}) of the $l=3000$ patchy kSZ, as a function of $z_r$ and $\Delta_z$. We leave blank under-sampled bins of ($z_r$, $\Delta_z$), defined as those for which the variance of the mean is larger than the mean of the variance.

We note that our estimates of $\bar{\mathcal{D}}^{\rm pkSZ}_{3000}$ are tens of percent higher compared to some recent estimates, for a given combination of $z_r$ and $\Delta_z$ \citep{Gorce2022, Chen2022}.  This might be in part due to different sampling of EoR morphologies, or to differences in how the patchy kSZ power is defined. Indeed, we estimate a difference of $\Delta z=1$ on the lower redshift bound of the integral in equation~\ref{general_equation} to result in a $\sim 0.2\,\mu\mathrm{K}^2$ difference in the patchy kSZ amplitude. Another potential source of disagreement could stem from using the Limber approximation to compute the patchy kSZ spectrum from the power spectrum of the density-weighted peculiar velocity field of the free electrons (e.g. \citealt{Ma2002, Gorce2020, Paul2021, Choudhury2021}), rather than ray-tracing the signal.
However, we find a good agreement between the two approaches on the scales of interest ($l \gtrsim 1000$).
We also compute the slope of the $\bar{\mathcal{D}}^{\rm pkSZ}_{3000}$ - $\Delta_z$ relation using the full posterior sample, finding values that are roughly $15 \%$ larger than \citet{Battaglia2013} and \citet{Chen2022}.  A more detailed comparison with other analysis is not possible given the differences in modelling, definitions, and EoR parameters, and would require a dedicated study.

In the left panel we also show the marginalized 1D PDFs of $z_r$ ({\it top}) and $\Delta_z$ ({\it right}).  Our {\it with kSZ} posterior corresponds to the following constraints on the EoR history parameters: $z_r = 7.12_{-0.41}^{+0.44}$ and $\Delta_z = 1.16_{-0.19}^{+0.24}$ (68\% C.I.).  As noted in previous studies, there is a strong degeneracy between $z_r$ and $\Delta_z$, as either a later or a shorter EoR decreases the patchy kSZ power.  Our median recovered values of $\Delta_z$ are consistent with those from other recent analyses of the SPT observation, including \citet{Reichardt2021} who found $\Delta_z = 1.1^{+1.6}_{-0.7}$ (68\% C.I.) and \citet{Choudhury2021} who found $\Delta_z = 1.30^{+0.19}_{-0.60}$.  Our limits are however $\sim 3$ times tighter compared to \citet{Reichardt2021} since we use additional, complementary observations in the likelihood.

In the right panel of Fig.~\ref{mean_sigma} we quantify the scatter in the $l=3000$ patchy kSZ power, {\it at fixed values of $z_r$ and $\Delta_z$}.  We see that the r.m.s. scatter in the power is generally at the level of a few percent.  Thus varying the $l=3000$ kSZ power amplitude at a function of only $z_r$ and $\Delta_z$, without considering the EoR morphology, cannot yield an accuracy on $\bar{\mathcal{D}}^{\rm pkSZ}_{3000}$ better than $\sim$ few percent.  
We stress also that this is a conservative estimate, since we only compute the scatter in the kSZ power for our relatively narrow {\it with kSZ} posterior.  Studies that do not consider complementary EoR and galaxy observations in the likelihood would result in broader posteriors with correspondingly larger scatter in the mean power.  Indeed, by sampling a broader range of models, \citet{Paul2021} find a larger r.m.s. scatter, of order $\sim 0.4 \mu$K$^2$ for the kSZ power at a fixed $z_r$ and $\Delta_z$.

It is important to note that without using complimentary observations in the likelihood, both the distributions of ($z_r$, $\Delta_z$) seen in the left panel and the scatter in the kSZ power at a fixed EoR history seen in the right panel would be considerably broader.  As noted earlier, the current SPT detection is low S/N and by itself is very not constraining.  Only in combination with complimentary observations can we obtain tight constraints on the EoR history and not be sensitive to our choice of priors.

\section{Conclusions}
\label{conclusion}

The patchy kSZ signal is an integral probe of the timing and morphology of the EoR. Recently, \citet{Reichardt2021} have claimed a detection of the patchy kSZ signal ($\mathcal{D}_{3000}^{\textrm{pkSZ}} = 1.1_{-0.7}^{+1.0} \,\mu\textrm{K}^2$). In the future, we expect a dramatic increase in S/N from telescopes such as CMB-S4 and Simons Observatory enhancing the potential of using kSZ measurements for EoR science.


In this work we quantify what we can learn about the EoR from the patchy kSZ signal. We modify the public {\tt 21cmFAST} code to produce forward-models of the patchy kSZ signal.  We then perform Bayesian inference by sampling galaxy properties and using the recent kSZ measurement together with other observations in the likelihood. These include: (i) high-$z$ UV LFs; (ii) Ly$\alpha$ forest opacity distributions; (iii) the Lyman forest pixel dark fraction and (iv) CMB optical depth. 


In order to quantify the additional constraining power of the patchy kSZ we computed two posteriors: one based on \citet{Qin2021} (using datasets (i)--(iv); \textit{without kSZ}) and one with an additional likelihood term for the recent measurement of the patchy kSZ cited above (\textit{with kSZ}). We found that the addition of the kSZ measurement shifts the posterior distribution in favor of faster and later reionization models (Fig.~\ref{corner_plot}). This results in a lower optical depth to the CMB: $\tau_e = 0.052_{-0.008}^{+0.009}$ (68\% C.I.).  

The shift to later and more rapid EoR implies a lower ionizing escape fraction with a very weak positive scaling with halo mass. 
 The average $f_{\rm esc}$ of typical galaxies driving the EoR is a few percent.  We disfavor a strong evolution of $f_{\rm esc}$ with galaxy mass.

We also present constraints on common empirical parameters characterizing the midpoint and duration of reionization, respectively $z_r = 7.10_{-0.41}^{+0.44}$ and $\Delta_z = 1.16_{-0.19}^{+0.24}$ (68\% C.I.), consistent with other recent results \citep{Reichardt2021, Choudhury2021}.
We show that the scatter in patchy kSZ power at $l=3000$, {\it at a fixed  $z_r$ and $\Delta_z$}, is of order $\sim$ few percent.  Thus the interpretation of current kSZ data can be done using only these two summary statistics.  However, without a physical model it would be difficult to assign prior probabilities or use complimentary observations in the likelihood.  

Future observations should further improve the measurement of the patchy kSZ signal \citep{Abazajian2016}.
 To forecast the resulting improvement in parameter constraints, we also create a mock observation with the measurement error reduced to $0.1 \mu K ^2$, centered on the MAP model from our inference.  Such a futuristic observation can reduce the uncertainties on the recovered EoR history by factors of $\sim$2, 3.  However, if the patchy kSZ power is confirmed to be low ($\mathcal{D}_{3000}^{\textrm{pkSZ}} \lesssim 2 \,\mu\textrm{K}^2$), it would result in a mild tension with the CMB $\tau_e$ inferred from primary CMB anisotropies.

\section*{Acknowledgements}

We thank C. Mason and A. Ferrara for insightful comments on a draft version of this manuscript.  We gratefully acknowledge computational resources of the Center for High Performance Computing (CHPC) at Scuola Normale Superiore (SNS). A.M. acknowledges support from the Ministry of Universities and Research (MUR) through the PNRR project "Centro Nazionale di Ricerca in High Performance Computing, Big Data e Quantum Computing" and the PRO3 project "Data Science methods for Multi-Messenger Astrophysics and Cosmology". Y.Q. acknowledges that part of this work was supported by the Australian Research Council Centre of Excellence for All Sky Astrophysics in 3 Dimensions (ASTRO 3D), through project \#CE170100013. A.G.'s work is supported by the McGill Astrophysics Fellowship funded by the Trottier Chair in Astrophysics, as well as the Canadian Institute for Advanced Research (CIFAR) Azrieli Global Scholars program and the Canada 150 Programme. 

\section*{Data Availability}

The data underlying this article will be shared on reasonable request to the corresponding author. The code will be made public by merging into the main ${\tt 21cmFAST}$ branch, after the manuscript is accepted for publication.



\bibliographystyle{mnras}
\bibliography{ksz_paper} 




\appendix

\section{Calibrating simulations to account for missing large-scale kSZ power}
\label{appendixA}

As discussed in Section \ref{estimating_section}, large boxes are required to accurately simulate the patchy kSZ signal. 
\citet{Shaw2012} find that a simulation box of side length $100 \,h^{-1}\textrm{Mpc}$ would miss about $60 \%$ of the kSZ power.
However, using large simulations which also resolve small-scale physics in forward modeling is computationally impractical. Although one could account for missing large-scale power analytically \citep{Park2013, Gorce2020}, such perturbative approaches are approximate and have only been tested with a few models. Instead, here we compute the kSZ signal directly from multiple, smaller-box realizations of the signal and statistically characterize the missing power comparing to a large-box realization \citep[see also][]{Iliev2007}.

\begin{figure}
    \includegraphics[width=1.0\columnwidth]{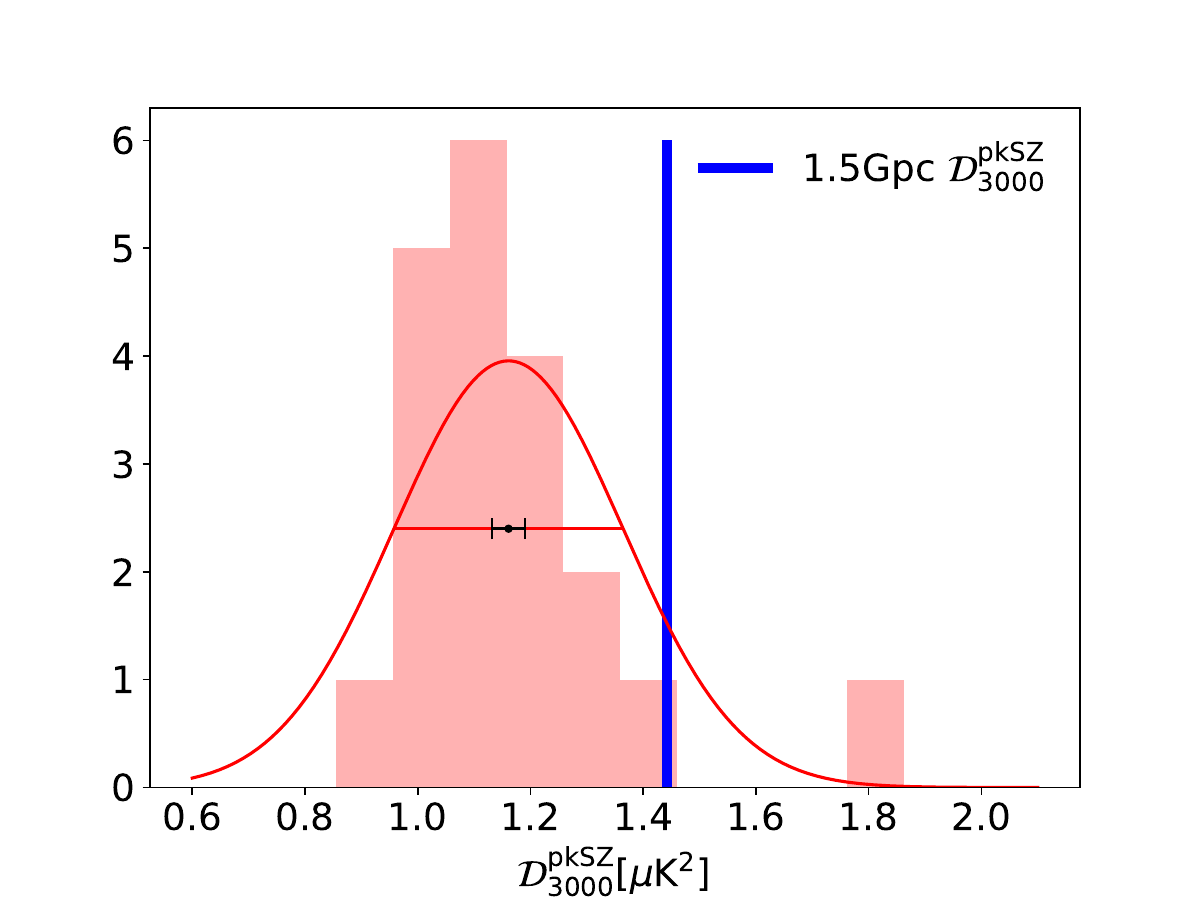}
    \caption{Histogram of the $l=3000$ patchy kSZ amplitudes generated from $500$\,Mpc boxes by varying the cosmic initial seed (see text for details). The solid red line corresponds to a Gaussian fit to the histogram, with the standard error $\sigma$ denoted as a solid line.  The vertical blue line denotes the value for the $1.5$\,Gpc box.   We see that the small boxes on average underestimate the kSZ power at $l=3000$ by $\sim20\%$, though with sizable scatter.  For illustrative purposes, we also demarcate with the black segment the size of the  1 $\sigma$ Poisson uncertainty on the mean power, arising from sampling the power spectrum with a limited number of modes for a 500 Mpc box.  We see that the cosmic variance from varying the seed is much larger than the Poison sample variance.}
    \label{scale_factors}
\end{figure}

We pick a random sample from the posterior distribution of \citet{Qin2021}, corresponding to the {\it without kSZ} posterior. For this set of astrophysical parameters, we compute the patchy kSZ power using a 1.5\,Gpc simulation, run on a 1050$^3$ grid.  We then generate 20 realizations of the smaller-box simulations used in our inference (500\,Mpc on a $256^3$ grid), using the same astrophysical parameters but varying the initial random seed.  When constructing the lightcones using the 500\,Mpc simulations, we rotate the coeval boxes to minimize duplication of structures due to periodic boundary conditions (e.g. \citealt{Mesinger2012}). We also performed a resolution check and found a negligible difference in the kSZ power with respect to the resolution. The resulting histogram of $\mathcal{D}^{\rm pkSZ}_{3000}$ from the 500 Mpc simulations is shown in Fig. \ref{scale_factors}, together with the value from the 1.5\,Gpc simulation (blue vertical line).  We compute the ratio of the missing power as $f_{0.5\textrm{Gpc}}^{1.5\textrm{Gpc}} = \mathcal{D}_{3000}^{\textrm{pkSZ} - 1.5\textrm{Gpc}} / \mathcal{D}_{3000}^{\textrm{pkSZ} - 500\textrm{Mpc}}$.  Using these 20 realizations, we find $f_{0.5\textrm{Gpc}}^{1.5\textrm{Gpc}} = 1.27 \pm 0.19$.   We include this scaling factor and associated uncertainty in the likelihood when performing inference:
\begin{equation}
    \ln \mathcal{L}_{\rm kSZ} = -\frac{1}{2} \left(\frac{\mathcal{D}_{3000}^{\textrm{kSZ,SPT}} - \mathcal{D}_{3000}^{\textrm{kSZ,model}}}{\sigma_a + \sigma_b  \left(\mathcal{D}_{3000}^{\textrm{kSZ,SPT}} - \mathcal{D}_{3000}^{\textrm{kSZ,mock}} \right)}\right)^2,
    \label{eq:likelihood}
\end{equation}
with $\sigma_a = 2\frac{\sigma_{u} \sigma_l}{\sigma_u + \sigma_l}$, $\sigma_b = \frac{\sigma_{u} - \sigma_{l}}{\sigma_{u} + \sigma_{l}}$. Here, $\sigma_{u}$ and $\sigma_{l}$ are upper and lower $68\%$ C.I. limits of the measurement. Since we are adding the scaling factor uncertainty in the quadrature, the final expressions for $\sigma_u$ and $\sigma_l$ are $\sigma_{u} = \sqrt{\sigma_{u, SPT}^2 + \sigma_{f_{0.5\textrm{Gpc}}^{1.5\textrm{Gpc}}}^2}$ and $\sigma_{l} = \sqrt{\sigma_{l, SPT}^2 + \sigma_{f_{0.5\textrm{Gpc}}^{1.5\textrm{Gpc}}}^2}$, where the measurement is expressed as $(\mathcal{D}_{3000}^{\textrm{kSZ,SPT}})_{-\sigma_{l,SPT}}^{+\sigma_{u,SPT}} = 1.1_{-0.7}^{+1.0} \mu \textrm{K}^2$. Log-likelihood written in equation~(\ref{eq:likelihood}) is a  Gaussian whose width depends on the parameter value and it's used for the asymmetric statistical errors of the measurement \citep[see e.g.][]{Barlow2004}.

\begin{table*}
 \caption{Astrophysical parameters used for the scaling test. $f_{0.5\textrm{Gpc}}^{1.5\textrm{Gpc}}$ is the power spectrum scaling factor (see text). Note that the parameter combination 2 is used in Fig.~\ref{kSZ plot}.}
 \label{tab_four_params}
 \begin{tabular}{lccccccc|c}
  \hline
   & $\log_{10} (f_\ast )$ & $\alpha_\ast$ & $\log_{10} (f_{\textrm{esc}} )$ & $\alpha_{\textrm{esc}}$ & $\log_{10} (M_{\textrm{turn}} /M_\odot )$ &$t_\ast$ & $\mathcal{D}_{3000}^{\rm pkSZ}$[$\mu$K$^2$] & $f_{0.5\textrm{Gpc}}^{1.5\textrm{Gpc}}$ \\
  \hline
  1 & -1.437 & 0.559 & -1.239 & 0.093 & 8.515 & 0.332 & 1.522 & 1.176\\
  2 & -1.416 & 0.614 & -1.780 & 0.474 & 8.622 & 0.392 & 1.076 &1.132\\
  3 & -1.498 & 0.493 & -1.201 & 0.175 & 8.668 & 0.282 & 1.469 & 1.132\\
  4 & -1.144 & 0.477 & -1.577 & 0.209 & 8.787 & 0.591 & 1.442 & 1.136\\
  \hline
 \end{tabular}
\end{table*}

Note that since we are varying the seed, the variance in $f_{0.5\textrm{Gpc}}^{1.5\textrm{Gpc}}$ also includes the Poisson uncertainty on the mean power, stemming from the fact that the power is estimated from a finite number of wavemodes. The later is illustrated as a solid black segment in Fig. \ref{scale_factors}, and has a subdominant contribution to the scatter in $\mathcal{D}^\mathrm{pkSZ}_{3000}$ from the 500\,Mpc simulations.

How much does the scaling factor, $f_{0.5\textrm{Gpc}}^{1.5\textrm{Gpc}}$, depend on the choice of astrophysical parameters?  Unfortunately, it would be computationally impractical to repeat the above calibration procedure over our entire 6D astrophysical parameter space.  Instead we sample four different astrophysical parameters from the {\it without kSZ} posterior, and compute $f_{0.5\textrm{Gpc}}^{1.5\textrm{Gpc}}$ using a single 1.5\,Gpc and 500\,Mpc simulation for each parameter set.  The parameters and corresponding scale factors are listed in the Table~\ref{tab_four_params}.  Reassuringly the scaling factors vary by only $\sim 3 \%$ between the four parameter combinations.  This is much smaller than the $\sim 0.7 \mu \textrm{K}^2$ measurement uncertainty, justifying our assumption of a constant $f_{0.5\textrm{Gpc}}^{1.5\textrm{Gpc}}$.

\section{Likelihoods used for inference}
\label{appendixB}

In Section~\ref{sec:complimentary_observations} we introduce various EoR observations that we used to perform inference of astrophysical parameters ($\theta$). Here we write out likelihoods for those observations, used in Section~\ref{sec:inference}:

\begin{itemize}
    \item{Lyman $\alpha$ forest opacity distributions - The log likelihood for one redshift, $z$, and effective optical depth, $\tau_{\rm eff}$, bin is assumed to be Gaussian:
    \begin{equation}
    \label{eq:lyalpha-lnl}
    \ln \mathcal{L}_{\alpha, z, \tau_{\rm eff}} (\theta) = - 0.5 X^{\rm T} \Sigma^{-1} X,
    \end{equation} where $\rm X$ is the difference between the model and the observed effective optical depth PDF in that bin and $\Sigma$ is the total error covariance matrix. More details about the covariance matrix can be found in appendices of \citet{Qin2021}.  The total log-likelihood for the Lyman $\alpha$ forest is the sum over redshift and optical depth bins:
    \begin{equation}
        \ln \mathcal{L}_{\alpha}(\theta) = \sum_{z} \sum_{\tau_{\rm eff}} \ln \mathcal{L}_{\alpha, z, \tau_{\rm eff}}(\theta)
    \end{equation} for $z\in \{5.4,5.6,5.8,6.0\}$.} and $\tau_{\rm eff} < 8$.
    \item{Dark fraction in the Ly $\alpha$ and Ly$\beta$ forests - the log likelihood is given as:
    \begin{equation}
    \ln \mathcal{L}_{\rm DF} (\theta)= \begin{cases}
    0 & \text{if $\overline{x}_{\rm HI, z}(\theta) \leq 0.06$} \\
    - \frac{1}{2} \frac{\left(\overline{x}_{\rm HI, z}(\theta) - 0.06\right)^2}{\sigma_{\rm DF}^2} & \text{otherwise}
    \end{cases}
    \end{equation} where $\overline{x}_{\rm HI,z}$ is the modelled neutral fraction at $z=5.9$ and $\sigma_{\rm DF} = 0.05$.}
    \item{High-redshift galaxy UV luminosity functions (UV LFs) - The log likelihood is the sum over redshifts and magnitudes given by \citet{Bouwens2015, Bouwens2016} and \citet{Oesch18}:
    \begin{equation}
        \ln \mathcal{L}_{\rm LF}(\theta) = -0.5 \sum_z \sum_{M_{\rm UV}} \left(\frac{\phi_{\rm LF, model}(\theta, z, M_{\rm UV}) - \phi_{\rm LF, obs}(z, M_{\rm UV})}{\sigma_{\rm LF}(z, M_{\rm UV})} \right)^2.
    \end{equation}
    Here $\phi_{\rm LF, model}$ is the modelled luminosity function at a given redshift and magnitude and $\phi_{\rm LF, obs}$ is the observed one, with the corresponding uncertainties, $\sigma_{\rm LF}$. Summation over redshifts is done for $z \in \{6,7,8,10\}$.}
    \item{The CMB optical depth - The log likelihood for the CMB optical depth is given as:
    \begin{equation}
        \ln \mathcal{L}_{\tau_{\rm e}} (\theta)= - \frac{1}{2} \left( \frac{\tau_{\rm e, model}(\theta) - \tau_{ \rm e, obs}}{\sigma_{\tau_{\rm e}}} \right)^2,
    \end{equation}}
    where $\tau_{\rm e, model}$ is the modelled CMB optical depth and $\tau_{\rm e, obs} = 0.0561$ is the observed one from \citet{Planck2020} with $\sigma_{\tau_{\rm e}} = 0.071$ the corresponding uncertainty. 
\end{itemize}
\bsp	
\label{lastpage}
\end{document}